# A Multi-Agent-Based Rolling Optimization Method for Restoration Scheduling of Electrical Distribution Systems with Distributed Generation

Donghan Feng, *Senior Member, IEEE*, Fan Wu, *Student Member, IEEE*, Yun Zhou, *Member, IEEE*, Usama Rahman, Xiaojin Zhao, Chen Fang

*Abstract*—Resilience against major disasters is the most essential characteristic of future electrical distribution systems (EDS). A multi-agent-based rolling optimization method for EDS restoration scheduling is proposed in this paper. When a blackout occurs, considering the risk of losing the centralized authority due to the failure of the common core communication network, the agents available after disasters or cyber-attacks identify the communication-connected parts (CCPs) in the EDS with distributed communication. A multi-time interval optimization model is formulated and solved by the agents for the restoration scheduling of a CCP. A rolling optimization process for the entire EDS restoration is proposed. During the scheduling/rescheduling in the rolling process, the CCPs in the EDS are reidentified and the restoration schedules for the CCPs are updated. Through decentralized decision-making and rolling optimization, EDS restoration scheduling can automatically start and periodically update itself, providing effective solutions for EDS restoration scheduling in a blackout event. A modified IEEE 123-bus EDS is utilized to demonstrate the effectiveness of the proposed method.

## I. INTRODUCTION

As the utility electrical infrastructure ages and the demand for electricity continues to increase, the impacts of major interruptions of the electrical infrastructure will be more intense [1]. Resilience against major disasters (e.g., floods, hurricanes and earthquakes) are considered by the U.S. Department of Energy (DOE) as the most essential characteristic of future electrical distribution systems (EDS) [2].

In recent worldwide outage events, after a major disaster or cyber-attack, it has taken hours or even days to restore the entire grid, e.g., 6 hours for the Ukraine blackout (Dec. 23th, 2015) [3] and 50 hours for the South Australia blackout (Sep. 28th, 2016) [4]. In the early restoration stage, if utility power from the transmission grid is unavailable, the EDS could consist of a multi-microgrid system made up of several self-healing microgrid islands, each equipped with distributed generators (DGs) and energy storages (ESs) to provide a balancing service to a wider area during blackouts.

Intelligent algorithms, such as genetic algorithms [5] and artificial neural networks [6], have been used in EDS restoration for a long time. Nevertheless, while these methods are proficient in providing an "acceptable" result under complex restoration requirements, they cannot guarantee the global optimality of the solution theoretically. Fuzzy set-based approaches [7], [8] can provide proper assessments of restoration goals. However, the restoration results are highly dependent on the design of the fuzzy sets, which is based on subjective experience rather than objective estimation. Using a stochastic method to model the uncertainty in an EDS restoration process [9] cannot deal with supplemental resources and the changing topology of the EDS. The solution of a stochastic model resorts to a Monte Carlo simulation, which makes this method difficult to use online.

The EDS restoration scheduling problem analysis in this paper aims to maximize the total prioritized loads restored in the EDS by forming microgrid islands powered by local distributed generation, including DGs and ESs, in the early restoration stage after a blackout. In [10], DGs, ESs, and electric vehicles (EVs) are considered to provide a continuous and stable power supply to the power loads, and a restoration strategy for the isolated EDS after the blackout is proposed. In [11], taking into account the uncertainty of the load and output of the DGs, a robust restoration decision-making model for the EDS is established. Considering the microgrid availability, hierarchical restoration schedules can be adopted to restore the critical loads of the EDS [12], [13]. The EDS restoration

---

D. Feng, F. Wu, Y. Zhou (corresponding author), U. Rahman, and X. Zhao are with the Key Laboratory of Control of Power Transmission and Conversion of the Ministry of Education, Department of Electrical Engineering, Shanghai Jiao Tong University, Shanghai, CO 200240 China (e-mail: seed@sjtu.edu.cn; nanflas.will@sjtu.edu.cn; yun.zhou@sjtu.edu.cn; usama.rehan21@sjtu.edu.cn; zhaoxiaojin@sjtu.edu.cn).

C. Fang is with the Electric Power Research Institute, State Grid Shanghai Municipal Electric Power Company, Shanghai, CO 200437 China (email: fangc02@gmail.com).

scheduling methods proposed in [10]-[13] are based on centralized optimization models. However, considering the risk of losing the centralized authority for restoration scheduling in the EDS due to physical damage or core communication network failures caused by major disasters or cyber-attacks, decentralized decision-making methods for EDS restoration scheduling should be developed.

A decentralized decision-making environment (e.g., a multi-agent system (MAS)) with two-way distributed communication for restoration scheduling can enhance the resilience of the EDS in a blackout event. In [14], considering only the local communication available, a distributed multi-agent-based method is proposed to achieve an optimal microgrids formation schedule for the EDS to restore the critical loads from a power outage. In [10], a distributed multi-agent-based load restoration algorithm for a microgrid is proposed, in which the stability (convergence) of a consensus algorithm is rigorously discussed. An agent-based consensus algorithm is adopted in [14] and [15] for global information discovery. Though the time-variant topology of the communication network is considered in both articles, the number of all agents in the MAS is fixed and known to each agent. However, this requirement cannot always be satisfied after a disaster when the MAS is broken into several connected parts and is inconsistent with the time-variant nature of the restoration process, such that reenergized, repaired and temporarily built agents in the previous restoration duration will join the MAS gradually. The fixed-number agent-based consensus algorithms in [14] and [15] need alteration to resolve this issue.

EDS restoration scheduling is a multi-time interval optimization problem. The dynamic system state (e.g., newly discovered DGs/ESs, reenergized or repaired agents, repaired buses/feeders) and possible deviations between the actual staged restoration results and the optimized schedules will affect the validity of the optimized restoration schedules in the subsequent restoration intervals.

The EDS restoration schedules need periodic revision to tackle these uncertainties. The rolling optimization method is an effective way of dealing with uncertainty in planning or scheduling. The main idea of 'rolling' is rescheduling periodically according to the updated information. In [16] and [17], different rolling-horizon approaches are introduced to deal with the uncertainty associated with renewable energy resources, load consumption and other parameters in an operation optimization.

In this paper, a multi-agent-based rolling optimization method for restoration scheduling of the EDS is proposed. A decentralized MAS with two-way distributed communication is proposed for EDS restoration scheduling. In the rolling optimization process, at the first scheduling moment, the communication-connected parts (CCPs) in the EDS are first identified by the information discovery process (IDP) of the available agents in the EDS at the beginning of the rolling optimization horizon. Second, based on the information discovered by the IDP, the restoration schedule for each CCP is determined by agents through solving the multi-time interval restoration scheduling optimization model. In the following rescheduling moments, the CCPs in the EDS are reidentified by all available agents, and the restoration schedule of the EDS is updated.

To the best of our knowledge, this method is the first that applies rolling optimization to EDS restoration scheduling. In the proposed method, through a decentralized online rolling decision-making procedure, the EDS restoration scheduling can automatically start and periodically update itself to adjust to the time-varying system states and uncertainties. The introduced definite MILP model can provide effective solutions for EDS restoration scheduling in a blackout event. The rest of this paper is organized as follows. The general multi-time interval optimization model for restoration scheduling is presented in Section II. The multi-agent-based rolling optimization method for EDS restoration scheduling is described in Section III. Case studies to validate the proposed restoration scheduling method are included in Section IV. Eventually, Section V concludes the paper.

## II. Restoration Scheduling Optimization Model

The general multi-time interval restoration scheduling optimization model for the CCP identified in the EDS is formulated in this section. The advanced restoration control period (i.e., the rolling optimization horizon) considered for the restoration scheduling optimization model is denoted by $T$, the discrete time step is denoted by $\Delta t$, and the number of time intervals in $T$ is given by $T/\Delta t$.

### A. Objective Function

The main objective of the restoration scheduling problem is to maximize the restored load in the given restoration period [11].

$$\max \sum_{n \in N_T} \sum_{i \in S_{\text{bus}}^{t_c}} (\omega_{L1} P_{L1,i}^{t_c+n\Delta t} + \omega_{L2} P_{L2,i}^{t_c+n\Delta t} + \omega_{L3} P_{L3,i}^{t_c+n\Delta t}) \Delta t \quad (1)$$

The detailed objective function is shown in (1), where $t_c$ is the beginning moment of the advance restoration control period, $N_T = \{0,1,...,T/\Delta t - 1\}$ is the set of discrete time intervals in the advance control period, $S_{\text{bus}}^{t_c}$ is the set of buses in the use state or that can be restored to the use state in $[t_c, t_c + T]$ in the CCP, $P_{L1,i}^{t_c+n\Delta t}$, $P_{L2,i}^{t_c+n\Delta t}$ and $P_{L3,i}^{t_c+n\Delta t}$ are the first-, second- and third-class active loads restored at bus $i$ at the discrete time moment $t_c + n\Delta t$, respectively, and $\omega_{L1}$, $\omega_{L2}$ and $\omega_{L3}$ are weighting coefficients of the first-, second- and third-class loads accordingly.

### B. Constraints

The constraints considered in the optimization model are presented in this part, including the network power flow constraints, the radial topology constraints, the bus power generation and load constraints, and other essential constraints.

*1) Network Power Flow Constraints*

Considering the advanced control period and based on the Distflow method, the network power flow constraints of the radial distribution power network are formulated in (2)-(9).

$$\sum_{ki \in S_{\text{feeder}}^{t_c}} P_{ki}^{t_c+n\Delta t} - \sum_{ij \in S_{\text{feeder}}^{t_c}} (P_{ij}^{t_c+n\Delta t} + r_{ij} I_{\text{sqr},ij}^{t_c+n\Delta t}) + \\ P_{G,i}^{t_c+n\Delta t} - P_{L,i}^{t_c+n\Delta t} = 0 \quad i \in S_{\text{bus}}^{t_c}, n \in N_T \quad (2)$$

$$\sum_{ki \in S_{\text{feeder}}^{t_c}} Q_{ki}^{t_c+n\Delta t} - \sum_{ij \in S_{\text{feeder}}^{t_c}} (Q_{ij}^{t_c+n\Delta t} + x_{ij} I_{\text{sqr},ij}^{t_c+n\Delta t}) + \\ Q_{G,i}^{t_c+n\Delta t} - Q_{L,i}^{t_c+n\Delta t} = 0 \quad i \in S_{\text{bus}}^{t_c}, n \in N_T \quad (3)$$

$$V_{\text{sqr},i}^{t_c+n\Delta t} - V_{\text{sqr},j}^{t_c+n\Delta t} = 2(P_{ij}^{t_c+n\Delta t} r_{ij} + Q_{ij}^{t_c+n\Delta t} x_{ij}) \\ + (r_{ij}^2 + x_{ij}^2) I_{\text{sqr},ij}^{t_c+n\Delta t} \quad w_{ij}^{t_c+n\Delta t} = 1, ij \in S_{\text{feeder}}^{t_c}, n \in N_T \quad (4)$$

$$(V_{\text{bus}}^{\text{norm}})^2 I_{\text{sqr},ij}^{t_c+n\Delta t} = f(P_{ij}^{t_c+n\Delta t}, r_{ij}(I_{\max,ij})^2, \Lambda) \\ + f(Q_{ij}^{t_c+n\Delta t}, x_{ij}(I_{\max,ij})^2, \Lambda) \quad ij \in S_{\text{feeder}}^{t_c}, n \in N_T \quad (5)$$

$$-w_{ij}^{t_c+n\Delta t} r_{ij}(I_{\max,ij})^2 \leq P_{ij}^{t_c+n\Delta t} \leq w_{ij}^{t_c+n\Delta t} r_{ij}(I_{\max,ij})^2 \\ ij \in S_{\text{feeder}}^{t_c}, n \in N_T \quad (6)$$

$$-w_{ij}^{t_c+n\Delta t} x_{ij}(I_{\max,ij})^2 \leq Q_{ij}^{t_c+n\Delta t} \leq w_{ij}^{t_c+n\Delta t} x_{ij}(I_{\max,ij})^2 \\ ij \in S_{\text{feeder}}^{t_c}, n \in N_T \quad (7)$$

$$(V_{\text{bus}}^{\min})^2 \leq V_{\text{sqr},i}^{t_c+n\Delta t} \leq (V_{\text{bus}}^{\max})^2 \quad v_i^{t_c+n\Delta t} = 1, i \in S_{\text{bus}}^{t_c}, n \in N_T \quad (8)$$

$$0 \leq I_{\text{sqr},ij}^{t_c+n\Delta t} \leq (I_{\max,ij})^2 \quad w_{ij}^{t_c+n\Delta t} = 1, ij \in S_{\text{feeder}}^{t_c}, n \in N_T \quad (9)$$

Constraints (2) and (3) are the active and reactive power balance equations. $P_{G,i}^{t_c+n\Delta t}/Q_{G,i}^{t_c+n\Delta t}$ is the active/reactive generation power of bus $i$ at $t_c+n\Delta t$ in $[t_c, t_c+T]$. $P_{L,i}^{t_c+n\Delta t}/Q_{L,i}^{t_c+n\Delta t}$ is the active/reactive restored load of bus $i$. $S_{\text{feeder}}^{t_c}$ is the set of feeders in the use state or that can be restored to the use state in $[t_c, t_c+T]$ in the CCP. $P_{ij}^{t_c+n\Delta t}/Q_{ij}^{t_c+n\Delta t}$ is the active/reactive power of feeder $ij$. $r_{ij}/x_{ij}$ is the resistance/reactance of feeder $ij$. $I_{\text{sqr},ij}^{t_c+n\Delta t}$ is the quadratic terms of the magnitude of the feeder current.

The voltage difference across feeder $ij$ can be obtained from (4), where $V_{\text{sqr},i}^{t_c+n\Delta t}$ is the quadratic term of the magnitude of the bus voltage. The binary variable $w_{ij}^{t_c+n\Delta t}$ is the state variable of feeder $ij$ (i.e., $w_{ij}^{t_c+n\Delta t}=1$ for the in use state and $w_{ij}^{t_c+n\Delta t}=0$ for the not in use state). Notice that for the condition constraint in (4) and later in this section, by introducing a sufficiently large constant $M$, the condition constraint can be transformed into two regular linear constraints.

Constraint (5) is a linear simplification of the equation $V_{\text{sqr},i}^{t_c+n\Delta t} I_{\text{sqr},ij}^{t_c+n\Delta t} = (P_{ij}^{t_c+n\Delta t})^2 + (Q_{ij}^{t_c+n\Delta t})^2$, while the detailed simplification process and definition of the self-optimal piecewise linear (PWL) approximation function $f(y, \bar{y}, \Lambda)$ can be referred to in [18]. $V_{\text{bus}}^{\text{norm}}$ is the nominal bus voltage magnitude. $I_{\max,ij}$ is the upper bound of the current magnitude of feeder $ij$. And $\Lambda$ is the number of discretizations used in the PWL function.

Constraints (6)-(9) are the upper and lower bound limits of $P_{ij}^{t_c+n\Delta t}$, $Q_{ij}^{t_c+n\Delta t}$, $V_{\text{sqr},i}^{t_c+n\Delta t}$ and $I_{\text{sqr},ij}^{t_c+n\Delta t}$, respectively. Similarly to $w_{ij}^{t_c+n\Delta t}$, $v_i^{t_c+n\Delta t}$ is the binary state variable of bus $i$. $V_{\text{bus}}^{\max}/V_{\text{bus}}^{\min}$ is the upper/lower bound of the bus voltage magnitude.

In addition, for $v_i^{t_c+n\Delta t}$ and $w_{ij}^{t_c+n\Delta t}$ in (2)-(9) and later in this section, if the use state of a bus or feeder is non-optional for some or all intervals in $[t_c, t_c+T]$, the corresponding $v_i^{t_c+n\Delta t}$ or $w_{ij}^{t_c+n\Delta t}$ should be set to fixed values. Note that the constraints in (2)-(9) are for a balanced EDS and can also be extended for a three-phase unbalanced EDS [19].

*2) Radial Topology Constraints*

Considering that the number of microgrid islands in the CCP is indeterminate and changeable during different discrete time intervals in the restoration process (e.g., through the merging of existed islands), the radial topology constraints proposed in [11] and [20] may be inapplicable.

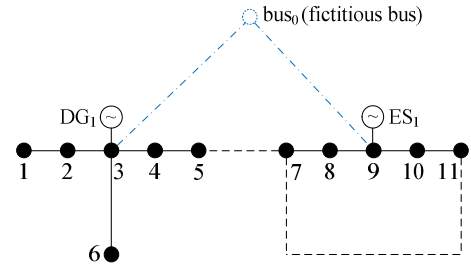

Fig. 1. Directed fictitious network of an EDS with DGs and ESs.

$$S_{\text{bus\_f}}^{t_c} = S_{\text{bus}}^{t_c} + \{0\} \quad (10)$$

$$S_{\text{feeder\_f}}^{t_c} = S_{\text{feeder}}^{t_c} + \{0j \mid j \in S_{\text{bus\_DG}}^{t_c} + S_{\text{bus\_ES}}^{t_c}\} \quad (11)$$

$$S_{\text{feeder\_f\_dir}}^{t_c} = \{ij \mid ij \in S_{\text{feeder\_f}}^{t_c} \text{ or } ji \in S_{\text{feeder\_f}}^{t_c}\} \quad (12)$$

On the basis of the idea of the fictitious network and fictitious power, directed fictitious network-based radial topology constraints can be suitable for a CCP with an indeterminate number of islands [21]. The schematic diagram of the directed fictitious network of an EDS is shown in Fig. 1. A fictitious bus ($bus_0$ in Fig. 1) with fictitious feeders connecting the fictitious bus and all DG buses and ES buses is added to the original network and makes up the fictitious network. In (11), $S_{\text{bus\_DG}}^{t_c}/S_{\text{bus\_ES}}^{t_c}$ is the set of DG/ES buses. Additionally, the sets of buses, feeders, and directed feeders ($S_{\text{bus\_f}}^{t_c}$, $S_{\text{feeder\_f}}^{t_c}$, and $S_{\text{feeder\_f\_dir}}^{t_c}$) of the directed fictitious network are defined in (10)-(12), respectively.

The radial constraints based on the directed fictitious network are presented in (13)-(20). The name of the fictitious bus ($bus_0$) is simplified as 0 for simplification in the formulas.

$$\sum_{ki \in S_{\text{feeder\_f\_dir}}^{t_c}} H_{ki}^{t_c+n\Delta t} - \sum_{ij \in S_{\text{feeder\_f\_dir}}^{t_c}} H_{ij}^{t_c+n\Delta t} = v_{f,i}^{t_c+n\Delta t} \\ i \in S_{\text{bus\_f}}^{t_c} - \{0\}, n \in N_T \quad (13)$$

$$0 \leq H_{ij}^{t_c+n\Delta t} \leq w_{f\_dir,ij}^{t_c+n\Delta t} \left| S_{\text{bus\_f}}^{t_c} \right| \quad ij \in S_{\text{feeder\_f}}^{t_c}, n \in N_T \quad (14)$$

$$v_{f,i}^{t_c+n\Delta t} + v_{f,j}^{t_c+n\Delta t} \geq 2(w_{f\_dir,ij}^{t_c+n\Delta t} + w_{f\_dir,ji}^{t_c+n\Delta t}) \quad ij \in S_{\text{feeder\_f}}^{t_c}, n \in N_T \quad (15)$$

$$v_{f,i}^{t_c+n\Delta t} = v_i^{t_c+n\Delta t} \quad i \in S_{bus}^{t_c}, n \in N_T \quad (16)$$

$$w_{f\_dir,ij}^{t_c+n\Delta t} + w_{f\_dir,ji}^{t_c+n\Delta t} = w_{ij}^{t_c+n\Delta t} \quad ij \in S_{feeder}^{t_c}, n \in N_T \quad (17)$$

$$v_{f,0}^{t_c+n\Delta t} = 1 \quad n \in N_T \quad (18)$$

$$w_{f\_dir,j0}^{t_c+n\Delta t} = 0 \quad j0 \in S_{feeder\_f\_dir}^{t_c}, n \in N_T \quad (19)$$

$$\sum_{ki \in S_{feeder\_f\_dir}^{t_c}} w_{f\_dir,ki}^{t_c+n\Delta t} = v_{f,i}^{t_c+n\Delta t} \quad i \in S_{bus\_f}^{t_c} - \{0\}, n \in N_T \quad (20)$$

In (13), $v_{f,i}^{t_c+n\Delta t}$ is the binary state variable of bus $i$ in the fictitious network and $H_{ij}^{t_c+n\Delta t}$ is the fictitious power of directed feeder $ij$ in the fictitious network. In (14), $w_{f\_dir,ij}^{t_c+n\Delta t}$ is the binary state variable of feeder $ij$ in the fictitious network. From (13) and (14), every bus (except bus$_0$) in the use state in the fictitious network has a unit of fictitious load. All fictitious power transmitted in the fictitious network is generated from bus$_0$. The fictitious power balance equation in (13) ensures the connection of the fictitious network (all buses and feeders in the use state).

The constraints of the state variables of buses and feeders in the directed fictitious network are described in (15)-(19). The relationship between the bus and feeder state variables is indicated in (15). The mapping relationships of the state variables in the fictitious network and the original network are represented in (16) and (17). The default values of bus$_0$ and fictitious feeders pointed to bus$_0$ are set in (18) and (19), respectively.

For a connected network including all buses and feeders in the use state, constraint (20) makes every bus (except bus$_0$) in the use state have only one parent bus, which ensures that the connected network is a treelike branching network. From constraints (13)-(20), the radial topology of the fictitious network can be guaranteed, and accordingly, the radial topology of all microgrid islands in the CCP (the original network) can be guaranteed.

3) *Bus Power Generation and Load Constraints*

The bus power generation and load constraints in the multi-time interval optimization model are formulated in (21)-(45) in this part.

$$P_{G,i}^{t_c+n\Delta t} = 0 \quad i \in S_{bus}^{t_c} - S_{bus\_DG}^{t_c} - S_{bus\_ES}^{t_c}, n \in N_T \quad (21)$$

$$Q_{G,i}^{t_c+n\Delta t} = 0 \quad i \in S_{bus}^{t_c} - S_{bus\_DG}^{t_c} - S_{bus\_ES}^{t_c}, n \in N_T \quad (22)$$

In (21) and (22), $P_{G,i}^{t_c+n\Delta t}$ and $Q_{G,i}^{t_c+n\Delta t}$ are limited to zero for buses without DG or ES, respectively.

$$P_{G,i}^{t_c+n\Delta t} = P_{DG,i}^{t_c+n\Delta t} \quad i \in S_{bus\_DG}^{t_c}, n \in N_T \quad (23)$$

$$Q_{G,i}^{t_c+n\Delta t} = Q_{DG,i}^{t_c+n\Delta t} \quad i \in S_{bus\_DG}^{t_c}, n \in N_T \quad (24)$$

$$v_i^{t_c+n\Delta t} P_{DG,i}^{min} \leq P_{DG,i}^{t_c+n\Delta t} \leq v_i^{t_c+n\Delta t} P_{DG,i}^{max} \quad i \in S_{bus\_DG}^{t_c}, n \in N_T \quad (25)$$

$$P_{DG,i}^{t_c+n\Delta t} = 0 \quad t_c + n\Delta t < t_{DG\_start,i}^{t_c} + T_{DG,i}^{syn}, i \in S_{bus\_DG}^{t_c}, n \in N_T \quad (26)$$

$$-v_i^{t_c+n\Delta t} P_{DG\_ramp,i}^{max} \leq P_{DG,i}^{t_c+(n+1)\Delta t} - P_{DG,i}^{t_c+n\Delta t} \leq v_i^{t_c+n\Delta t} P_{DG\_ramp,i}^{max}$$
$$t_c + n\Delta t \geq t_{DG\_start,i}^{t_c} + T_{DG,i}^{syn}, i \in S_{bus\_DG}^{t_c}, n \in N_T' \quad (27)$$

$$v_i^{t_c+n\Delta t} \frac{P_{DG,i}^{t_c+n\Delta t}}{P_{DG,i}^{max}} Q_{DG,i}^{min} \leq Q_{DG,i}^{t_c+n\Delta t} \leq v_i^{t_c+n\Delta t} \frac{P_{DG,i}^{t_c+n\Delta t}}{P_{DG,i}^{max}} Q_{DG,i}^{max}$$
$$i \in S_{bus\_DG}^{t_c}, n \in N_T \quad (28)$$

The power generation constraints for DG buses are presented in (23)-(28). In (23) and (24), $P_{DG,i}^{t_c+n\Delta t} / Q_{DG,i}^{t_c+n\Delta t}$ is the net active/reactive output power of the DG at bus $i$. In (25), $P_{DG,i}^{max} / P_{DG,i}^{min}$ is the upper/lower bound of $P_{DG,i}^{t_c+n\Delta t}$. The starting and ramping constraints are depicted in (26) and (27). Referring to a typical DG operating curve in [22], in (26), $t_{DG\_start,i}^{t_c}$ is the moment when the DG is ready to start up, and $T_{DG,i}^{syn}$ is the synchronization time parameter of the DG. In (27), $N_T' = \{0,1,...,T/\Delta t - 2\}$ and $P_{DG\_ramp,i}^{max}$ is the upper ramping active power limit parameter of the DG between two adjacent discrete time intervals. In (28), $Q_{DG,i}^{max} / Q_{DG,i}^{min}$ is the upper/lower bound of $Q_{DG,i}^{t_c+n\Delta t}$ when $P_{DG,i}^{t_c+n\Delta t} = P_{DG,i}^{max}$.

$$P_{G,i}^{t_c+n\Delta t} = P_{ES,i}^{t_c+n\Delta t} \quad i \in S_{bus\_ES}^{t_c}, n \in N_T \quad (29)$$

$$Q_{G,i}^{t_c+n\Delta t} = Q_{ES,i}^{t_c+n\Delta t} \quad i \in S_{bus\_ES}^{t_c}, n \in N_T \quad (30)$$

$$P_{ES,i}^{t_c+n\Delta t} = P_{ES\_ch,i}^{t_c+n\Delta t} - P_{ES\_dis,i}^{t_c+n\Delta t} \quad i \in S_{bus\_ES}^{t_c}, n \in N_T \quad (31)$$

$$0 \leq P_{ES\_ch,i}^{t_c+n\Delta t} \leq v_{ch,i}^{t_c+n\Delta t} P_{ES\_ch,i}^{max} \quad i \in S_{bus\_ES}^{t_c}, n \in N_T \quad (32)$$

$$0 \leq P_{ES\_dis,i}^{t_c+n\Delta t} \leq v_{dis,i}^{t_c+n\Delta t} P_{ES\_dis,i}^{max} \quad i \in S_{bus\_ES}^{t_c}, n \in N_T \quad (33)$$

$$v_{ch,i}^{t_c+n\Delta t} + v_{dis,i}^{t_c+n\Delta t} \leq v_i^{t_c+n\Delta t} \quad i \in S_{bus\_ES}^{t_c}, n \in N_T \quad (34)$$

$$Q_{ES\_min,i}^{t_c+n\Delta t} \leq Q_{ES,i}^{t_c+n\Delta t} \leq Q_{ES\_max,i}^{t_c+n\Delta t} \quad i \in S_{bus\_ES}^{t_c}, n \in N_T \quad (35)$$

$$SoC_{ES,i}^{t_c+(n+1)\Delta t} - SoC_{ES,i}^{t_c+n\Delta t} =$$
$$\frac{(P_{ES\_ch,i}^{t_c+n\Delta t}\eta_{ch,i} - P_{ES\_dis,i}^{t_c+n\Delta t}\eta_{dis,i})\Delta t}{C_{ES,i}} \quad i \in S_{bus\_ES}^{t_c}, n \in N_T' \quad (36)$$

$$SoC_{ES\_min,i} \leq SoC_{ES,i}^{t_c+n\Delta t} \leq SoC_{ES\_max,i} \quad i \in S_{bus\_ES}^{t_c}, n \in N_T \quad (37)$$

Considering the battery energy storage systems used, the generation power constraints for ES buses are presented in (29)-(37). In (29) and (30), $P_{ES,i}^{t_c+n\Delta t} / Q_{ES,i}^{t_c+n\Delta t}$ is the net active/reactive output power of the ES at bus $i$. In (31), $P_{ES\_ch,i}^{t_c+n\Delta t} / P_{ES\_dis,i}^{t_c+n\Delta t}$ is the active charge/discharge power of the ES. In (32) and (33), $P_{ES\_ch,i}^{max} / P_{ES\_dis,i}^{max}$ is the charge/discharge power upper bound and $v_{ch,i}^{t_c+n\Delta t} / v_{dis,i}^{t_c+n\Delta t}$ is the binary charge/discharge state variable. Constraint (34) will avoid the ES being operated in charging and discharging modes simultaneously. In (35), $Q_{ES\_max,i}^{t_c+n\Delta t} / Q_{ES\_min,i}^{t_c+n\Delta t}$ is the upper/lower bound of $Q_{ES,i}^{t_c+n\Delta t}$. According to Fig. 2 [23], the set values of $Q_{ES\_max,i}^{t_c+n\Delta t} / Q_{ES\_min,i}^{t_c+n\Delta t}$ can be simplified as shown in Table I. Constraint (36) represents the state of charge (SoC) variation of an ES, where $SoC_{ES,i}^{t_c+n\Delta t}$ is the SoC of the ES with the upper/lower bound $SoC_{ES\_max,i} / SoC_{ES\_min,i}$ in (37). $C_{ES,i}$ represents the rated energy capacity of the ES, and $\eta_{ch,i}/\eta_{dis,i}$ is the charge/discharge efficiency parameter.

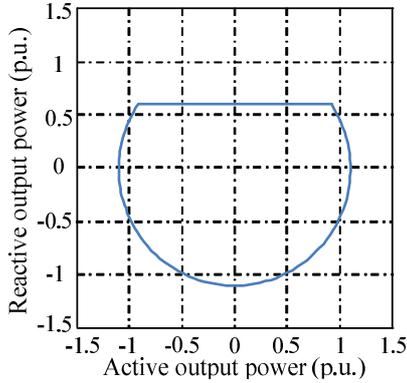

Fig. 2. Battery energy storage system reactive power capacity.

TABLE I
ES REACTIVE POWER ADJUSTABLE RANGE

| Range of $\left\|P_{ES,i}^{t_c+n\Delta t}\right\|$ (p.u.) | $[Q_{ES\_min,i}^{t_c+n\Delta t}, Q_{ES\_max,i}^{t_c+n\Delta t}]$ (p.u.) |
|---|---|
| [0,0.2] | [-1.1,0.6] |
| (0.2,0.4] | [-1,0.6] |
| (0.4,0.6] | [-0.9,0.6] |
| (0.6,0.8] | [-0.75,0.6] |
| (0.8,1] | [-0.5,0.5] |

$$P_{L,i}^{t_c+n\Delta t} = P_{L1,i}^{t_c+n\Delta t} + P_{L2,i}^{t_c+n\Delta t} + P_{L3,i}^{t_c+n\Delta t} \quad i \in S_{bus}^{t_c}, n \in N_T \quad (38)$$

$$0 \leq P_{L1,i}^{t_c+n\Delta t} \leq v_i^{t_c+n\Delta t} P_{L1\_par,i}^{t_c+n\Delta t} \quad i \in S_{bus}^{t_c}, n \in N_T \quad (39)$$

$$0 \leq P_{L2,i}^{t_c+n\Delta t} \leq v_i^{t_c+n\Delta t} P_{L2\_par,i}^{t_c+n\Delta t} \quad i \in S_{bus}^{t_c}, n \in N_T \quad (40)$$

$$0 \leq P_{L3,i}^{t_c+n\Delta t} \leq v_i^{t_c+n\Delta t} P_{L3\_par,i}^{t_c+n\Delta t} \quad i \in S_{bus}^{t_c}, n \in N_T \quad (41)$$

$$Q_{L,i}^{t_c+n\Delta t} = Q_{L1,i}^{t_c+n\Delta t} + Q_{L2,i}^{t_c+n\Delta t} + Q_{L3,i}^{t_c+n\Delta t} \quad i \in S_{bus}^{t_c}, n \in N_T \quad (42)$$

$$Q_{L1,i}^{t_c+n\Delta t} = \frac{P_{L1,i}^{t_c+n\Delta t}}{P_{L1\_par,i}^{t_c+n\Delta t}} Q_{L1\_par,i}^{t_c+n\Delta t} \quad i \in S_{bus}^{t_c}, n \in N_T \quad (43)$$

$$Q_{L2,i}^{t_c+n\Delta t} = \frac{P_{L2,i}^{t_c+n\Delta t}}{P_{L2\_par,i}^{t_c+n\Delta t}} Q_{L2\_par,i}^{t_c+n\Delta t} \quad i \in S_{bus}^{t_c}, n \in N_T \quad (44)$$

$$Q_{L3,i}^{t_c+n\Delta t} = \frac{P_{L3,i}^{t_c+n\Delta t}}{P_{L3\_par,i}^{t_c+n\Delta t}} Q_{L3\_par,i}^{t_c+n\Delta t} \quad i \in S_{bus}^{t_c}, n \in N_T \quad (45)$$

The load constraints in the optimization model are presented in (38)-(45). In (39)-(41), $P_{L1\_par,i}^{t_c+n\Delta t}$, $P_{L2\_par,i}^{t_c+n\Delta t}$ and $P_{L3\_par,i}^{t_c+n\Delta t}$ are the first-, second- and third-class active bus load parameters. In (42), $Q_{L1,i}^{t_c+n\Delta t}$, $Q_{L2,i}^{t_c+n\Delta t}$ and $Q_{L3,i}^{t_c+n\Delta t}$ are the first-, second- and third-class reactive loads restored at bus $i$. Constraints (43)-(45) ensure that active and reactive loads are restored in proportion, where $Q_{L1\_par,i}^{t_c+n\Delta t}$, $Q_{L2\_par,i}^{t_c+n\Delta t}$ and $Q_{L3\_par,i}^{t_c+n\Delta t}$ are the first-, second- and third-class reactive bus load parameters.

*4) Other Essential Constraints*

Other essential constraints in the optimization model are formulated in (46)-(52).

$$P_{L,i}^{t_c+n\Delta t} \geq v_i^{t_c+n\Delta t} \lambda_{min,i}(P_{L1\_par,i}^{t_c+n\Delta t} + P_{L2\_par,i}^{t_c+n\Delta t} + P_{L3\_par,i}^{t_c+n\Delta t}) \quad (46)$$
$$i \in S_{bus}^{t_c}, n \in N_T$$

$$P_{L1,i}^{t_c+n\Delta t} \leq P_{L1,i}^{t_c+(n+1)\Delta t} \quad i \in S_{bus}^{t_c}, n \in N_T' \quad (47)$$

$$P_{L2,i}^{t_c+n\Delta t} \leq P_{L2,i}^{t_c+(n+1)\Delta t} \quad i \in S_{bus}^{t_c}, n \in N_T' \quad (48)$$

$$v_i^{t_c+n\Delta t} \leq v_i^{t_c+(n+1)\Delta t} \quad i \in S_{bus}^{t_c}, n \in N_T' \quad (49)$$

$$w_{ij}^{t_c+n\Delta t} \leq w_{ij}^{t_c+(n+1)\Delta t} \quad ij \in S_{feeder}^{t_c}, n \in N_T' \quad (50)$$

$$\{v_i^{t_c}, w_{ij}^{t_c}\} = \{v_i^{t_c}, w_{ij}^{t_c}\}_{observed} \quad i \in S_{bus}^{t_c}, ij \in S_{feeder}^{t_c} \quad (51)$$

$$\{P_{G,i}^{t_c}, P_{L1,i}^{t_c}, P_{L2,i}^{t_c}, P_{L3,i}^{t_c}, SoC_{ES,i}^{t_c}\} = \{P_{G,i}^{t_c}, P_{L1,i}^{t_c}, P_{L2,i}^{t_c}, P_{L3,i}^{t_c}, SoC_{ES,i}^{t_c}\}_{observed} \quad i \in S_{bus}^{t_c} \quad (52)$$

Constraint (46) illustrates that if a bus is restored, to ensure the basic function of the bus, a minimum percentage $\lambda_{min,i}$ of the bus load needs be restored. The restoration continuity constraints are presented in (47)-(50). The first- and second-class loads restored in the current discrete time interval cannot be decreased in later intervals. Additionally, the buses and feeders in the use state in the current discrete time interval should maintain the use state in later time intervals. In (51) and (52), the boundary conditions of the multi-time interval optimization model, $v_i^{t_c}$, $w_{ij}^{t_c}$, $P_{G,i}^{t_c}$, $P_{L1,i}^{t_c}$, $P_{L2,i}^{t_c}$, $P_{L3,i}^{t_c}$, $SoC_{ES,i}^{t_c}$, etc., in the first time interval should be set to the observed values at the current control moment of the advanced restoration control period.

To summarize this section, the multi-time interval optimization model of the restoration scheduling problem of the CCP is formulated with the objective function in (1) and constraints in (2)-(9) and (13)-(52). As all the constraints are linear constraints with binary variables, the integrated multi-time interval optimization model is a mixed-integer linear programming (MILP) model and can be solved by commercial solvers effectively.

### III. MULTI-AGENT-BASED ROLLING OPTIMIZATION PROCESS FOR EDS RESTORATION SCHEDULING

#### A. Framework of the Multi-Agent System

In this paper, an agent in the MAS refers to an individual device or a group of devices which have the ability to acquire local information, control local devices, communicate with other agents, and retain/manipulate the data of the whole distribution system [14]. With a local emergency power source and two-way distributed communication (e.g., through an ad hoc network [24]), agents can continue to work after a disaster or cyber-attack. After a serious event, the surviving agents can participate in the restoration, while those that are unavailable can be reenergized or repaired and can join the restoration gradually.

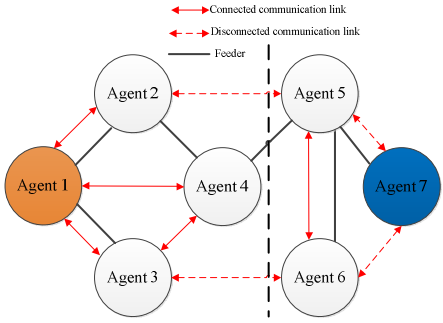

Fig. 3. MAS with seven agents.

The topology of the communication network may differ from that of the distribution system. From the sight of graph theory, an MAS can be regarded as a graph, an available agent can be considered as a node and a two-way communication link can be considered as an edge. A CCP is a maximally connected subgraph of the MAS. An example of an MAS with seven agents is shown in Fig. 3. Agent 1 is the agent with the computing ability for restoration scheduling. Agent 7 is supposed as unavailable, while the other agents are available. Accordingly, there exists two CCPs in Fig. 3, and Agents 1-4 form a CCP, while Agents 5 and 6 form another.

### B. Information Discovery Process (IDP)

In the following part, each bus is supposed to assign an agent. Each agent has the ability to control local devices and save the information of the whole distribution system. Only the information of the load, DGs, ESs, and states of the buses/feeders linked to this bus can be obtained directly. The other information needed for restoration scheduling can be acquired by communication with the agents nearby. As the distributed generations are the cores of the microgrid islands, the buses with DGs and ESs are assigned agents with the computing ability for restoration scheduling.

Referring to the multi-time interval optimization model formulated in Section II, the information stored in an agent at time $t_c$ must contain the following state information in a CCP.

1) Available states of buses: $A_{bus}^{t_c}$
2) Available states of feeders: $A_{feeder}^{t_c}$

If a bus or a feeder is available, its corresponding state is set at 1, whereas it is set 0 if it is not available. These two pieces of information are used to identify the network topology.

3) Parameters of feeders: $\{r_{ij}, x_{ij}\}$
4) Energized states of buses: $\{v_i^{t_c}\}$
5) Energized states of feeders: $\{w_{ij}^{t_c}\}$
6) Load restored states of buses: $\{P_{L1,i}^{t_c}, P_{L2,i}^{t_c}, P_{L3,i}^{t_c}\}$, and $\{Q_{L1,i}^{t_c}, Q_{L2,i}^{t_c}, Q_{L3,i}^{t_c}\}$.
7) State of DGs: $\{P_{DG,i}^{t_c}\}$ and $\{t_{DG\_start,i}^{t_c}\}$.
8) State of ESs: $\{SoC_{ES,i}^{t_c}\}$.

The other state information needed for restoration scheduling can be calculated during the scheduling process or prestored in the agent. All control variables are derived from the multi-time interval optimization. The agents that finish scheduling first can broadcast the values of control variables to the other agents in the same CCP.

In normal conditions, an MAS has a common communication network, and the agents within can communicate with each other with no limitations (all need for communication is satisfied). After a serious disaster, the common communication network could be destroyed, and the consistent power supply for an agent could be cut off simultaneously; thus, the communication of an agent would be supported by limited local power resources such that the communication ability of an agent would be limited. Considering the difficulty in rebuilding common communication and the limited communication ability of agents, a distributed method of communication has to be adopted for the IDP under this circumstance. The consensus algorithm is used in the IDP. The consensus of agents means that all agents in a CCP of the MAS reach an agreement about certain information of this CCP. With limited communication ability, each agent can still obtain the global information of this CCP through an appropriate consensus algorithm. Considering the latency of communication and the time consumption of the data manipulation process, a discrete-form consensus algorithm is preferred. In addition, since a disaster might change the configuration of the MAS, the consensus algorithm must be free of information about the topology of the whole system. The average consensus algorithm introduced in [25] is used in this paper.

$$X_i^{k+1} = X_i^k + \sum_{j \in N(i)} a_{ij}(X_j^k - X_i^k) \quad k=0,1,\cdots,k_{max} \quad (53)$$

In (53), $X_i^k$ is the abstract vector of the system parameters known to agent $i$ in the $k$th iteration, $X_i^0$ is the information originally owned by agent $i$, and all parameters referring to information acquired by other agents are all initiated to 0. $k_{max}$ is the predetermined maximum iteration number. $N(i)$ is the set of agents that have communication with agent $i$ (i.e., neighbors of agent $i$). The coefficient $a_{ij}$ is the Metropolis-Hasting weight calculated in (54), where $n_i$ is the number of neighbors of agent $i$.

$$a_{ij} = \begin{cases} 1/(\max\{n_i,n_j\}+1) & j \in N(i) \\ 1-\sum_{k \in N(i)} 1/(\max\{n_i,n_k\}+1) & j=i \\ 0 & j \notin N(i) \end{cases} \quad (54)$$

The asymptotic result of (53) with the coefficients decided by (54) is:

$$\lim_{k \to \infty} X_i^k = (1/N_{a,C})\sum_{j=1}^{N_{a,C}} X_j^0 \quad (55)$$

In (55), $N_{a,C}$ is the total number of agents in the CCP. The convergence of the above consensus algorithm can be proven by referring to [15]. It should be emphasized that the validity of applying (53) to the MAS depends on the assumption that every communication link in the MAS is two-way. Though the choice of $a_{ij}$ may influence the speed of convergence, the design of $a_{ij}$ is beyond the scope of this paper.

In some cases, the MAS may be divided into several CCPs after a disaster. Moreover, the recovery of communication is a dynamic process, which makes the topology of each CCP time-variant. Without the knowledge of $N_{a,C}$ in a CCP, the result of the average consensus algorithm in (55) would be meaningless since only an average result is obtained. To handle this problem, an indicator set with a dynamic size is assigned to every available agent. For agent $i$, the indicator set $I_i$ is initiated as $I_i^0 = \{(i,1)\}$, where 1 is the initial indicator corresponding to agent $i$. Before implementing the consensus algorithm on other values, (53) is applied to the indicator set. Supposing that in the iteration $k+1$, agent $i$ receives the information of agent $j$ (which may not be directly from agent $j$) for the first time, then $I_i^k$ will be initiated as $I_i^k = I_i^k \cup \{(j,0)\}$ before (53) is implemented. The indicator corresponding to agent $i$ will achieve $1/N_{a,C}$ asymptotically for all agents in the CCP. Then, the value of all parameters needed for scheduling the restoration will be initiated as $N_{a,C} X_i^0$.

By implementing the average consensus algorithm, the information known to any agent will asymptotically approach to the average among each agent in a CCP, which means that all information of this CCP is known to all agents inside it. A certain CCP has no information about other CCPs. After the synchronous IDP is complete, the restoration scheduling for the CCPs will be activated inside each CCP independently.

*C. Rolling Optimization Process of EDS Restoration Scheduling*

In general, the rolling optimization method transforms a stochastic process into a series of deterministic scenarios. Each scenario lasts for a period of time called a horizon. The horizon can be classified into two types: the prediction horizon and the control horizon. A control schedule for the whole prediction horizon is made based on the current information at the beginning of the prediction horizon and will be implemented at the beginning of the control horizon. The prediction horizon and the corresponding control horizon have the same start time. The control horizon is always shorter than the prediction horizon. The end of a control horizon is the beginning of the next prediction horizon.

Different rolling strategies are utilized in [16] and [17]. In [16], the lengths of the prediction horizon and control horizon are fixed, and both horizons move forward simultaneously. In [17], from the start to the end of a whole scheduling period, the control horizon is fixed, while the prediction horizon decreases gradually from the length of the whole scheduling period to the length of a control horizon. Since the finishing time of the restoration process can hardly be determined, a rolling strategy similar to that introduced in [16] is adopted in this paper.

A number of basic assumptions are made to support the rolling optimization process of the EDS restoration scheduling.

1) For each communication link in the multi-agent system, the communication between the two agents is two-way. This ensures that the applied consensus algorithm is always convergent.
2) Each available agent has universal time information (e.g., from the GPS timing system). All the agents in the whole distribution system are synchronized at the moment they begin to participate in the IDP. The IDP and restoration process are triggered by a universal clock.
3) The IDP of each agent starts at the same time and is completed before scheduling the restoration. If an agent resumes communication with other agents during or after an IDP, it cannot join this IDP and has to wait until the next IDP begins. In this circumstance, if the agent is associated with a resource, though communication with other agents is unavailable, the DG can independently start synchronization at once.
4) Once an agent participates in the IDP, it will not disconnect during this IDP.
5) The restoration schedules made for different agents in the same CCP are identical. The agent that first finishes the scheduling calculation will broadcast the result to all other agents in the CCP.

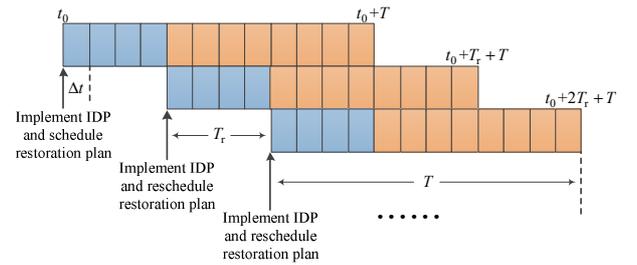

Fig. 4. Timeline of the rolling optimized process of EDS restoration scheduling.

Controlled by a universal clock, Fig. 4 shows the timeline of the rolling optimized process of EDS restoration scheduling, where $t_0$ is the restoration start moment of the EDS, $T$ is the rolling optimization horizon (i.e., the prediction horizon), $T_r$ is the rescheduling time gap between two adjacent optimization horizons (i.e., the control horizon), and $\Delta t$ is the discrete time step in the restoration scheduling optimization model for the CCPs.

The behavior of the agent on a bus with restoration scheduling ability is shown in Fig. 5 in detail. The agent first implements the IDP described in Section III.B and then models the optimization problem described in Section II based on the discovered information and solves the problem. The scheduling/rescheduling results will be broadcasted to all agents in the CCP. The other agents have similar behaviors to that shown in Fig. 5, excluding the CCP restoration scheduling step.

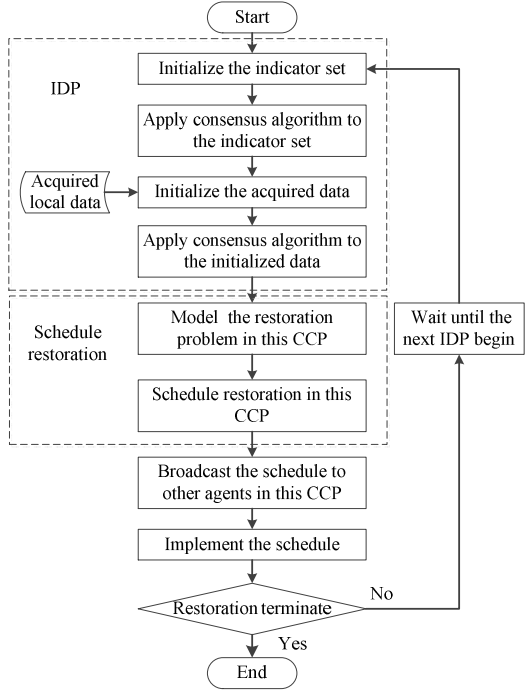

Fig. 5. Behaviors of the agent with the computing ability for restoration scheduling.

## IV. CASE STUDY

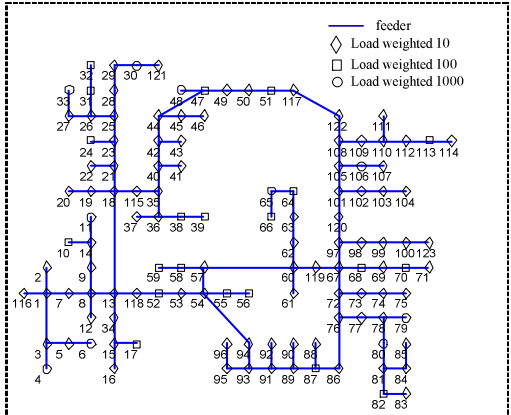

Fig. 6. Topology of the modified IEEE 123-bus EDS in the normal state.

The proposed multi-agent-based rolling optimization method for restoration scheduling of the EDS is tested using a modified IEEE 123-bus EDS. The network topology of the test system is shown in Fig. 6, while the original parameters of this system are obtained from [26]. The weights of the first-, second- and third-class of active loads in the test system ($\omega_{L1}$, $\omega_{L2}$, and $\omega_{L3}$) are set as 1000, 100, and 10, respectively. The communication network is supposed to have the same topology as the distribution network, and each bus agent can only communicate with the bus agent adjacent to it. In the rolling optimization process of EDS restoration scheduling shown in Fig. 4, the rolling optimization horizon $T$ is set to 120 min, and the rescheduling time gap $T_r$ is set to 30 min. The discrete time step $\Delta t$ in the restoration scheduling optimization model for the CCP is set to 5 min. For the IDP, the convergence condition of the consensus algorithm is set as the difference between two iterations, at less than $10^{-10}$. The proposed multi-agent-based rolling optimization process for EDS restoration scheduling is implemented in MATLAB with YALMIP, and Gurobi is used as the MILP solver.

### A. Restoration Scheduling Results under $T_r$ =30 min

With the restoration rolling optimization horizon of $T$ =120 min, the rescheduling time gap $T_r$ is set to 30 min, and the restoration scheduling is started at $t_c$ =0 min and updated at $t_c$ =30 min, 60 min, 90 min, etc. The restoration scheduling of the test system is completed through the multi-agent-based rolling optimization process proposed in Section III. The detailed restoration results at four sequential restoration scheduling (start) and rescheduling moments (i.e., $t_c$ =0 min, 30 min, 60 min, and 90 min) are summarized and analyzed in this part.

In the figures depicting the system, the different shapes of the buses represent different priorities. A dotted blue line between two bus agents means that there is a communication link between the two agents and that the feeder between the two buses is available. A solid green line means that the feeder is energized. All DGs are represented as red circles, while the red star represents the ES. The shade of the bus symbol corresponds to the percentage of the load restored at the bus. The darker the shade, the higher is the percentage of the load restored at the bus. For the bus with an ES, the shade of the star symbol illustrates the percentage of energy remaining. In the figures depicting the performance of the consensus algorithm, changes in the inverse number of the indicator of each agent are illustrated.

#### 1) Restoration Results at $t_c$ =0 min

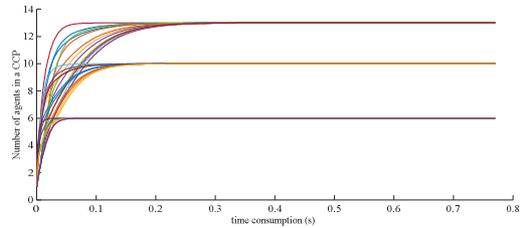

Fig. 7. Performance of the average consensus algorithm in identifying CCPs at the first scheduling moment $t_c$ =0 min in $T$ =[0 min, 120 min].

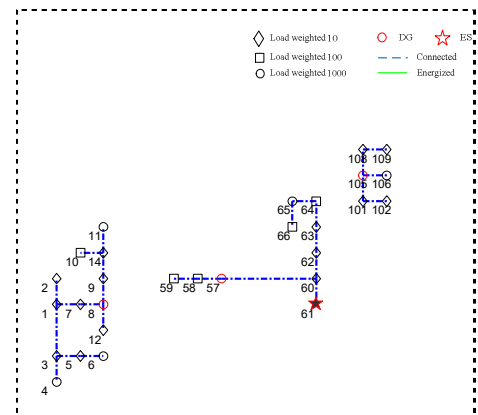

Fig. 8. CCPs identified and restoration results at $t_c$ =0 min.

The restoration of the test system is supposed to be started at the $t_c$ =0 min moment. Fig. 7 and 8 show the IDP results of the test system at $t_c$ =0 min. According to Fig. 7, there are 3 CCPs in the EDS, and each has 13, 10 and 6 agents. Fig. 8 shows the agents and connections available at $t_c$ =0 min. In total, 3 DGs and 1 ES are discovered and available for load restoration in the EDS. The major parameters of the DGs and ES are listed in Table II and III. $t^{t_c}_{DG\_start}$ of the DG on bus 8 is negative, which means that it is started 10 min before it is discovered at $t_c$ =0 min by the IDP.

TABLE II
PARAMETERS OF THE DISCOVERED ES AT BUS 61 IN THE TEST EDS AT $t_c$ =0 MIN

| $C_{ES}$ (kWh) | $P^{max}_{ES\_ch}$ (kW) | $P^{max}_{ES\_dis}$ (kW) | $Q_{max}$ (kVar) | $\eta_{ch}/\eta_{dis}$ | $SoC_{ES\_max}$ | $SoC_{ES\_min}$ | $SoC^0_{ES}$ |
|---|---|---|---|---|---|---|---|
| 200 | 50 | 50 | 50 | 0.85/1.15 | 0.95 | 0.05 | 0.8 |

TABLE III
PARAMETERS OF THE DISCOVERED DGs IN THE TEST EDS AT $t_c$ =0 MIN

| Bus | $P_{max}$ (kW) | $P_{min}$ (kW) | $Q_{max}$ (kVar) | $P^{max}_{DG\_ramp}$ (kW/min) | $T^{syn}_{DG}$ (min) | $t^{t_c}_{DG\_start}$ (min) |
|---|---|---|---|---|---|---|
| 8 | 200 | 33.3 | 150 | 11.1 | 10 | -10 |
| 57 | 300 | 66.7 | 200 | 16.7 | 15 | 0 |
| 105 | 200 | 33.3 | 150 | 11.1 | 10 | 0 |

TABLE IV
DGs AND ES ACTIVE POWER OUTPUTS AND DIFFERENT CLASS LOAD RESTORATION RESULTS AT $t_c$ =0 MIN

| $t_c$ | $\sum_{i\in S^{t_c}_{bus}} P^{t_c}_{G,i}$ (kW) | $\sum_{i\in S^{t_c}_{bus}} P^{t_c}_{L1,i}$ (kW) | $\sum_{i\in S^{t_c}_{bus}} P^{t_c}_{L2,i}$ (kW) | $\sum_{i\in S^{t_c}_{bus}} P^{t_c}_{L3,i}$ (kW) |
|---|---|---|---|---|
| 0 min | 0 | 0 | 0 | 0 |

Table IV tabulates the DGs and ES active power outputs and different class load restoration results at $t_c$ =0 min. As $t_c$ =0 min is the start of restoration, the sum of the DGs and ES active power outputs (second column in Table IV), the sum of first-class loads restored ($\omega_{L1}$ =1000, third column), the sum of second-class loads restored ($\omega_{L2}$ =100, fourth column), and the sum of third-class loads restored ($\omega_{L3}$ =10, fifth column) are all equal to zero. After the global information is obtained through the IDP, the corresponding multi-time interval restoration optimization model (formulated in (1), (2)-(9), and (13)-(52)) is solved by the agents with scheduling ability in each CCP; thus, the restoration schedule for the EDS in the following restoration rolling optimization horizon $T$ =[0 min, 120 min] is determined.

2) *Restoration Results at $t_c$ =30 min*

As $T_r$ is set to 30 min, the restoration schedules should be updated at $t_c$ =30 min. Table V summarizes the actual restoration results of the test system at $t_c$ =30 min according to the optimized EDS restoration schedules determined at $t_c$ =0 min. On the basis of the restoration results, referring to Fig. 5, the IDP is first implemented to update the system state information at $t_c$ =30 min.

TABLE V
DGs AND ES ACTIVE POWER OUTPUTS AND DIFFERENT CLASS LOAD RESTORATION RESULTS AT $t_c$ =30 MIN

| $t_c$ | $\sum_{i\in S^{t_c}_{bus}} P^{t_c}_{G,i}$ (kW) | $\sum_{i\in S^{t_c}_{bus}} P^{t_c}_{L1,i}$ (kW) | $\sum_{i\in S^{t_c}_{bus}} P^{t_c}_{L2,i}$ (kW) | $\sum_{i\in S^{t_c}_{bus}} P^{t_c}_{L3,i}$ (kW) |
|---|---|---|---|---|
| 30 min | 600.1 | 300.0 | 170.9 | 124.6 |

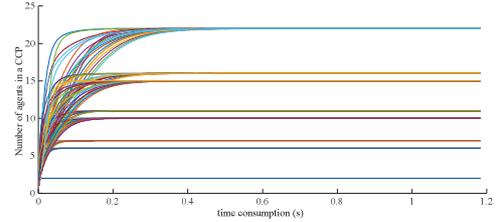

Fig. 9. Performance of the average consensus algorithm in identifying CCPs at the first scheduling moment $t_c$ =30 min in $T$ =[30 min, 150 min].

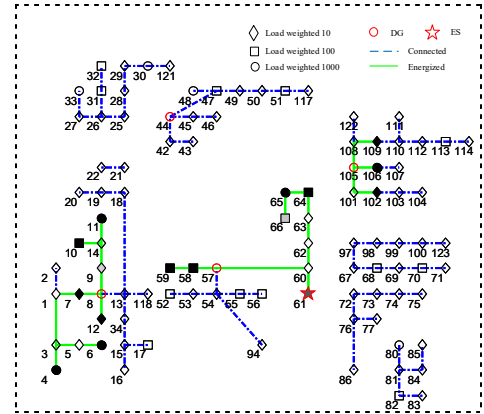

Fig. 10. CCPs identified and restoration results at $t_c$ =30 min.

TABLE VI
PARAMETERS OF THE NEWLY DISCOVERED DG IN THE TEST EDS AT $t_c$ =30 MIN

| Bus | $P_{max}$ (kW) | $P_{min}$ (kW) | $Q_{max}$ (kVar) | $P^{max}_{DG\_ramp}$ (kW/min) | $T^{syn}_{DG}$ (min) | $t^{t_c}_{DG\_start}$ (min) |
|---|---|---|---|---|---|---|
| 44 | 150 | 16.7 | 100 | 8.3 | 10 | 15 |

Fig. 9 and 10 show the IDP results of the test system at $t_c$ =30 min. The DG with the parameters in Table VI is newly discovered and available for load restoration in later restoration intervals. Based on the new information, the restoration schedule for the EDS in the following $T$ =[30 min, 150 min] is updated at $t_c$ =30 min.

3) *Restoration Results at $t_c$ =60 min*

Table VII summarizes the actual restoration results of the test system at $t_c$ =60 min. On the basis of the restoration results, the IDP is implemented to update the system state information at $t_c$ =60 min.

TABLE VII
DGs AND ES ACTIVE POWER OUTPUTS AND DIFFERENT CLASS LOAD RESTORATION RESULTS AT $t_c$ =60 MIN

| $t_c$ | $\sum_{i \in S_{bus}^{t_c}} P_{G,i}^{t_c}$ (kW) | $\sum_{i \in S_{bus}^{t_c}} P_{L1,i}^{t_c}$ (kW) | $\sum_{i \in S_{bus}^{t_c}} P_{L2,i}^{t_c}$ (kW) | $\sum_{i \in S_{bus}^{t_c}} P_{L3,i}^{t_c}$ (kW) |
|---|---|---|---|---|
| 60 min | 900.0 | 410.6 | 285.8 | 198.2 |

TABLE IX
DGs AND ES ACTIVE POWER OUTPUTS AND DIFFERENT CLASS LOAD RESTORATION RESULTS AT $t_c$ =90 MIN

| $t_c$ | $\sum_{i \in S_{bus}^{t_c}} P_{G,i}^{t_c}$ (kW) | $\sum_{i \in S_{bus}^{t_c}} P_{L1,i}^{t_c}$ (kW) | $\sum_{i \in S_{bus}^{t_c}} P_{L2,i}^{t_c}$ (kW) | $\sum_{i \in S_{bus}^{t_c}} P_{L3,i}^{t_c}$ (kW) |
|---|---|---|---|---|
| 90 min | 1227.1 | 670.0 | 503.8 | 47.3 |

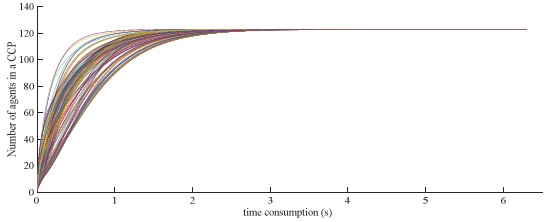

Fig. 11. Performance of the average consensus algorithm in identifying CCPs at the first scheduling moment $t_c$ =60 min in $T$ =[60 min, 180 min].

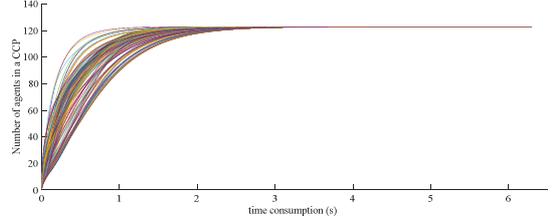

Fig. 13. Performance of the average consensus algorithm in identifying CCPs at the first scheduling moment $t_c$ =90 min in $T$ =[90 min, 210 min].

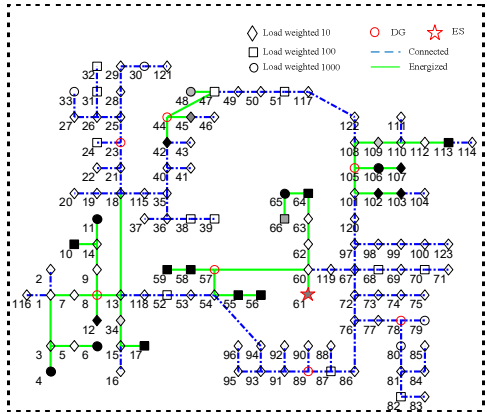

Fig. 12. CCPs identified and restoration results at $t_c$ =60 min.

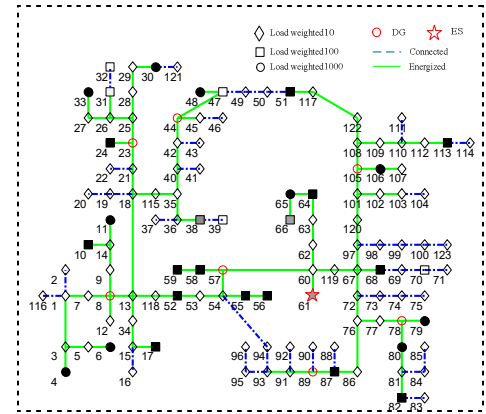

Fig. 14. CCPs identified and restoration results at $t_c$ =90 min.

TABLE VIII
PARAMETERS OF THE NEWLY DISCOVERED DG IN THE TEST EDS AT $t_c$ =60 MIN

| Bus | $P_{max}$ (kW) | $P_{min}$ (kW) | $Q_{max}$ (kVar) | $P_{DG\_ramp}^{max}$ (kW/min) | $T_{DG}^{syn}$ (min) | $t_{DG\_start}^{t_c}$ (min) |
|---|---|---|---|---|---|---|
| 23 | 100 | 5.6 | 60 | 5.6 | 5 | 40 |
| 78 | 120 | 13.3 | 100 | 6.7 | 10 | 50 |
| 89 | 120 | 13.3 | 100 | 6.7 | 10 | 20 |

TABLE X
DETAILED INFORMATION ABOUT THE CONSENSUS PROCESS AT DIFFERENT $t_c$

| $t_c$ | Iteration number | Time cost (s) |
|---|---|---|
| 0 min | 770 | 0.8 |
| 30 min | 1184 | 1.2 |
| 60 min | 6290 | 6.3 |
| 90 min | 6290 | 6.3 |

Fig. 11 and 12 show the IDP results of the test system at $t_c$ =60 min. The DGs with the parameters in Table VIII are newly discovered and available for load restoration in later restoration intervals.

4) *Restoration Results at $t_c$ =90 min*

Table IX summarizes the actual restoration results of the test system at $t_c$ =90 min. On the basis of the restoration results, the IDP is implemented to update the system state information at $t_c$ =90 min, and the corresponding results are shown in Fig. 13 and 14.

Referring to Fig. 4, the restoration schedules will be successively updated at $t_c$ =120 min, 150 min, 180 min, etc., unless the rolling optimization terminal condition is met (e.g., the power supply from the transmission network is recovered).

It is important to note that in the performance of the average consensus algorithm in identifying CCPs at different $t_c$ in Fig. 7, 9, 11, and 13, considering that the wireless communication latency is less than 1 ms when supported by next-generation communication technology [27], the time span of each iteration in (53) is set as 1 ms. In accordance with Fig. 7, 9, 11, and 13, the detailed iteration numbers and time costs of the consensus processes at different $t_c$ are listed in Table X. The time cost of the consensus processes makes up the majority of the IDP time span. Compared to $\Delta t$ and $T_r$ in the rolling optimization method, the time span for IDP can be neglected, which ensures the effectiveness of the proposed multi-agent-based rolling optimization method in EDS restoration scheduling.

### B. Restoration Scheduling Results under $T_r$ =45 min

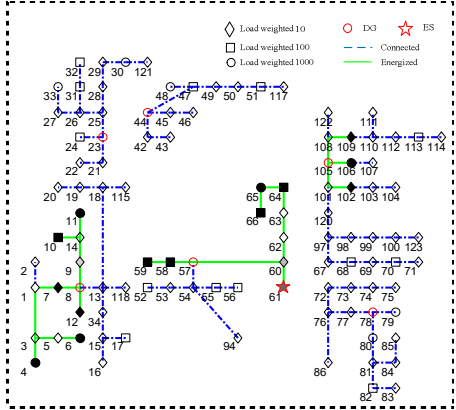

Fig. 15. CCPs identified and restoration results at $t_c$ =45 min under $T_r$ =45 min.

To further analyze the features of the proposed rolling optimization restoration scheduling method, the rescheduling time gap $T_r$ is set to 45 min in this part for comparison. The restoration results at $t_c$ =0 min under $T_r$ =45 min are the same as those in the $T_r$ =30 min case in the first subsection of Section IV.A. Similar to the rolling process in Section IV.A, the restoration results for the test system at $t_c$ =45 min and 90 min under $T_r$ =45 min are briefly shown in Fig. 15-16 and Tables XI-XIV.

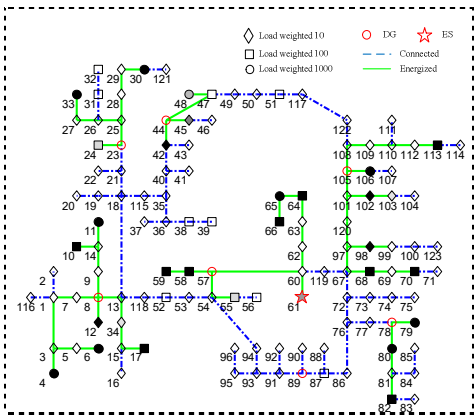

Fig. 16. CCPs identified and restoration results at $t_c$ =90 min under $T_r$ =45 min.

TABLE XI
DGs AND ES ACTIVE POWER OUTPUTS AND DIFFERENT CLASS LOAD RESTORATION RESULTS AT $t_c$ =45 MIN UNDER $T_r$ =45 MIN

| $t_c$ | $\sum_{i \in S^{t_c}_{bus}} P^{t_c}_{G,i}$ (kW) | $\sum_{i \in S^{t_c}_{bus}} P^{t_c}_{L1,i}$ (kW) | $\sum_{i \in S^{t_c}_{bus}} P^{t_c}_{L2,i}$ (kW) | $\sum_{i \in S^{t_c}_{bus}} P^{t_c}_{L3,i}$ (kW) |
|---|---|---|---|---|
| 45 min | 650.1 | 300 | 210.0 | 134.0 |

TABLE XII
PARAMETERS OF THE NEWLY DISCOVERED DGs IN THE TEST EDS AT $t_c$ =45 MIN UNDER $T_r$ =45 MIN

| Bus | $P_{max}$ (kW) | $P_{min}$ (kW) | $Q_{max}$ (kVar) | $P^{max}_{DG\_ramp}$ (kW/min) | $T^{syn}_{DG}$ (min) | $t^{t_c}_{DG\_start}$ (min) |
|---|---|---|---|---|---|---|
| 23 | 100 | 5.6 | 60 | 5.6 | 5 | 40 |
| 44 | 150 | 16.7 | 100 | 8.3 | 10 | 15 |
| 78 | 120 | 13.3 | 100 | 6.7 | 10 | 50 |

TABLE XIII
DGs AND ES ACTIVE POWER OUTPUTS AND DIFFERENT CLASS LOAD RESTORATION RESULTS AT $t_c$ =90 MIN UNDER $T_r$ =45 MIN

| $t_c$ | $\sum_{i \in S^{t_c}_{bus}} P^{t_c}_{G,i}$ (kW) | $\sum_{i \in S^{t_c}_{bus}} P^{t_c}_{L1,i}$ (kW) | $\sum_{i \in S^{t_c}_{bus}} P^{t_c}_{L2,i}$ (kW) | $\sum_{i \in S^{t_c}_{bus}} P^{t_c}_{L3,i}$ (kW) |
|---|---|---|---|---|
| 90 min | 1118.0 | 570.6 | 378.5 | 162.2 |

TABLE XIV
PARAMETERS OF THE NEWLY DISCOVERED DGs IN THE TEST EDS AT $t_c$ =90 MIN UNDER $T_r$ =45 MIN

| Bus | $P_{max}$ (kW) | $P_{min}$ (kW) | $Q_{max}$ (kVar) | $P^{max}_{DG\_ramp}$ (kW/min) | $T^{syn}_{DG}$ (min) | $t^{t_c}_{DG\_start}$ (min) |
|---|---|---|---|---|---|---|
| 89 | 120 | 13.3 | 100 | 6.7 | 10 | 20 |

The comparison of the load restoration results of the test system at $t_c$ =90 min under different set values of $T_r$ (30 min and 45 min) is shown in Table XV. At $t_c$ =90 min, the sum of the first-class loads restored (third column in Table XV), the sum of the second-class loads restored (fourth column), and the total loads restored (sixth column) in the load restoration results under $T_r$ =30 min are larger than those in the $T_r$ =45 min case, demonstrating a better restoration performance.

TABLE XV
LOAD RESTORATION RESULTS OF THE TEST SYSTEM AT $t_c$ =90 MIN UNDER $T_r$ =30 MIN AND $T_r$ =45 MIN

| $t_c$ | $T_r$ | $\sum_{i \in S^{t_c}_{bus}} P^{t_c}_{L1,i}$ (kW) | $\sum_{i \in S^{t_c}_{bus}} P^{t_c}_{L2,i}$ (kW) | $\sum_{i \in S^{t_c}_{bus}} P^{t_c}_{L3,i}$ (kW) | $\sum_{i \in S^{t_c}_{bus}} P^{t_c}_{L,i}$ (kW) |
|---|---|---|---|---|---|
| 90 min | 30 min | 670.0 | 503.8 | 47.3 | 1221.1 |
| 90 min | 45 min | 570.6 | 378.5 | 162.2 | 1111.3 |

Referring to Tables II, III, VI, and VIII in Section IV.A and Tables XII and XIV in this part, the rolling optimization restoration processes can consider the new restoration resources (e.g., newly discovered DGs/ESs, reenergized or repaired agents, repaired buses/feeders) during the whole restoration, which effectively enhances the restoration performance.

In general, a shorter $T_r$ leads to a better restoration performance. An adequate frequency of rescheduling should be considered in practical EDS restoration. Considering the set values of $\Delta t$ ($\Delta t$ =5 min or 10 min in practical restoration), frequent rescheduling ($T_r \leq$ 15 min) is unnecessary and will increase the pressure on the communication system. For the set value of $\Delta t$ =5 min in the case study, the set value of $T_r$ =30 min can yield a satisfying restoration performance of the test system.

## V. CONCLUSION

A multi-agent-based rolling optimization method for restoration scheduling of an EDS with distributed generation is proposed in this paper. Considering the risk of losing the centralized authority due to the failure of the core communication network, an MAS with distributed communication is introduced, and a rolling optimization process is newly established to realize decentralized decision-making for EDS restoration scheduling. Through decentralized decision-making and rolling optimization, EDS restoration scheduling can be automatically started and can periodically update itself, providing effective solutions for EDS restoration scheduling in a blackout event.

The proposed method is verified on a modified 123-bus EDS. In the case studies, the steps of the multi-agent-based rolling optimization method for EDS restoration scheduling are illustrated by detailed restoration results at four sequential restoration scheduling/rescheduling moments in the rolling process. Case studies demonstrate that the rolling restoration optimization process can consider new restoration resources to participate in the restoration in a timely manner and effectively enhance the restoration performance. To analyze the influence of the communication latency on the rolling restoration optimization process, the convergence and time cost of the average consensus algorithm for the IDP during restoration scheduling are studied. Comparisons between restoration scheduling results for the test system under different rescheduling time gaps are also included in the case studies to demonstrate the features of the proposed EDS restoration scheduling method.

From the perspective of resilience, the multi-agent-based restoration scheduling method can overcome the unreliability of centralized restoration scheduling after disasters. The rolling optimization process can consider new restoration resources in the EDS through periodically updating the restoration schedule. This paper mainly focuses on the load restoration of the EDS in the earlier restoration stages after a blackout. Future works will include the coordination of EDS restoration with transmission system restoration. The uncertainty of the distributed generation in EDS restoration will also be studied.

**Donghan Feng** received the B.S. and Ph.D. degrees from the Department of Electrical Engineering, Zhejiang University, Hangzhou, China, in 2003 and 2008, respectively.

Dr. Feng has been with the faculty of Shanghai Jiao Tong University (SJTU) since 2008, where he currently is a full professor. Dr. Feng also serves as the Deputy Director of the State Energy Smart Grid Research and Development Center, Shanghai, China. He was a Graduate Research Assistant with Tsinghua University from 2005 to 2006, a Visiting Scholar with the University of Hong Kong from 2006 to 2007, a Hans Christened Ørsted Postdoc with Technical University of Denmark, Lyngby, Denmark, from 2009 to 2010, a Visiting Research Scholar with the University of California, Berkely, USA, from 2015 to 2016.

He is the recipient of the Pujiang Scholar of Shanghai, the Morningstar Excellent Young Faculty Award of SJTU, and the Future Scientist Program of China Scholarship Council, Ministry of Education of China. He is a senior member of IEEE. His research interests include spot pricing in smart energy networks.

**Fan Wu** received the B.S. degree in electrical engineering from Shanghai Jiao Tong University, Shanghai, China, in 2017. Currently, he is pursuing the M.S. degree with the Key Laboratory of Control of Power Transmission and Conversion, Ministry of Education, Department of Electrical Engineering, Shanghai Jiao Tong University. His research interests include the power system optimization, hybrid renewable energy system and energy storage systems.

**Yun Zhou** received the B.S. (Hons.) and Ph.D. degrees in electrical engineering from Shanghai Jiao Tong University, Shanghai, China, in 2012 and 2017, respectively.

He participated in the joint Ph.D. program of Alstom Grid Technology Center Co. Ltd. and Shanghai Jiao Tong University, Shanghai, China, from Sep. 2013 to Jun. 2017. He was an R&D intern of Alstom Grid China Technology Center from Sep. 2013 to Oct. 2015, and was an R&D intern of GE Grid Solutions China Technology Center from Nov. 2015 to Jun. 2017. He has joined the faculty of the Electrical Engineering Department, Shanghai Jiao Tong University, Shanghai, China, as a Lecturer since Dec. 2017. His current research interests include power system restoration and energy internet.

**Usama Rahman** received the B.S. degree in electrical power engineering from the National University of Science Technology, Karachi, Pakistan, in 2013, and the M.S. degree in Electrical Power Systems from the North China Electric Power University, Beijing, China, in 2017. He is currently pursuing the Ph.D. in Power System Engineering from, Shanghai Jiao Tong University, Shanghai, China. He is also a part of faculty development programme from NUST, Pakistan. His research interest includes electrical power optimization, energy storage systems, artificial intelligence, hybrid renewable energy system and energy storage systems.

**Xiaojin Zhao** received the B.S. degree in electrical engineering from Shanghai Jiao Tong University, Shanghai, China, in 2019. Currently, she is pursuing the M.S. degree with the Key Laboratory of Control of Power Transmission and Conversion, Ministry of Education, Department of Electrical Engineering, Shanghai Jiao Tong University. Her current research interests include the modeling and optimization of the energy internet.

**Chen Fang** received the B.S. and Ph.D. degrees from the Department of Electrical Engineering, Tsinghua University, Beijing, China, in 2006 and 2011, respectively.

He is currently an Electrical Engineer with the Electric Power Research Institute, State Grid Shanghai Municipal Electric Power Company, Shanghai, China. His current research interests include renewable energy integration of smart grid and power storage technology.